\documentclass[hidelinks]{article}

\usepackage{arxiv}

\usepackage{url}
\usepackage{breakurl}
\usepackage{hyperref}
\usepackage{cite}
\usepackage{amsmath,amssymb,amsfonts}
\usepackage{algorithmic}
\usepackage{graphicx}
\usepackage{textcomp}
\usepackage{xcolor}
\usepackage{booktabs}
\usepackage{leftidx}
\usepackage{siunitx}
\usepackage{xspace}
\usepackage{paralist}
\usepackage{caption}
\usepackage{subcaption}
\usepackage{gnuplottex}
\definecolor{darkgreen}{rgb}{0.0, 0.5, 0.0}
\definecolor{amaranth}{rgb}{0.9, 0.17, 0.31}
\definecolor{azure}{rgb}{0.0, 0.5, 1.0}

\usepackage{listings}

\usepackage{pgfplots}
\pgfplotsset{compat=1.14}

\newcommand{\orcid}[1]{\href{https://orcid.org/#1}{\includegraphics[height=10pt]{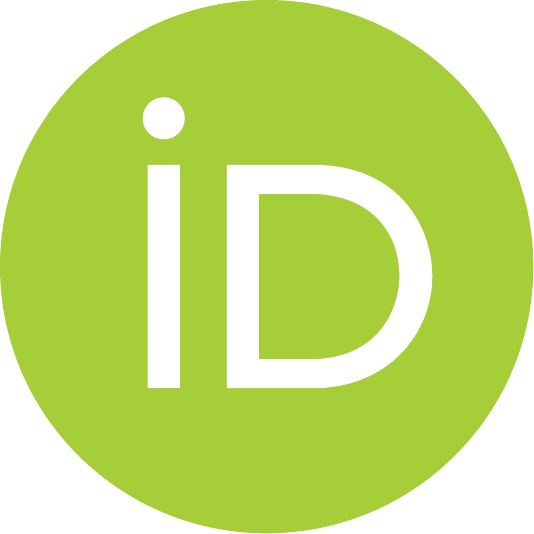}}}

\newcommand{\arm}{Arm\textsuperscript{\tiny\textregistered}}

\usepackage{todonotes}

\usepackage{xfrac}

\author{
    Nikunj Gupta\\
    Dept. of CSE, Indian Institute of Technology, Roorkee, India \\
    \textit{gnikunj@cct.lsu.edu}
    \And
    Steve R. Brandt, Bibek Wagle\orcid{0000-0001-6619-7115}, Nanmiao Wu, Alireza Kheirkhahan, Patrick Diehl\orcid{0000-0003-3922-8419}, Hartmut Kaiser\orcid{0000-0002-8712-2806}  \\
    Center of Computation \& Technology, Louisiana State University, Baton Rouge, LA, USA \\
    $\{$pdiehl,sbrandt,aliz,hkaiser$\}$@cct.lsu.edu, $\{$wnanmi1,bwagle3$\}$@lsu.edu
    \And
    Felix W. Baumann\orcid{0000-0001-5292-3040} \\
    PricewaterhouseCoopers GmbH Wirtschaftspr\"ufungsgesellschaft,
    Office of the Chief Data Officer,
    Frankfurt, Germany \\
    \textit{felix.baumann@pwc.com}
}

\begin{document}

\title{Deploying a Task-based Runtime System on Raspberry Pi Clusters}

\maketitle

\begin{abstract}
\arm~technology is becoming increasingly important in HPC.
Recently, Fugaku, an \arm-based system, was awarded the number one place in the Top500 list.
Raspberry Pis provide an inexpensive platform to become familiar with this architecture.
However, Pis can also be useful on their own. Here we describe our efforts to configure and benchmark the use of a Raspberry Pi cluster with the HPX/Phylanx platform (normally intended for use with HPC applications) and document the lessons we learned. First, we highlight the required changes in the configuration of the Pi to gain performance. Second, we explore how limited memory bandwidth limits the use of all cores in our shared memory benchmarks.
Third, we evaluate whether low network bandwidth affects distributed performance. Fourth, we discuss the power consumption and the resulting trade-off in cost of operation and performance.
\end{abstract}


\section{Introduction}
The {\arm} architecture is becoming increasingly important in the high-performance computing (HPC) and server world. Not only is the fastest super computer in the \textit{Top500}, Fugaku, an \arm-based system, but Sandia National Labs has announced the \arm-based Astra prototype cluster. One reason for the rising interest is energy efficiency, since it is one of the most significant costs for any supercomputer and a critical factor for building a petaflops cluster.

The Raspberry Pi is based on the {\arm} architecture and follows the {\arm} ISA. It has a single pipeline, Neon, to enable support for vectorization, which potentially increases its computing power by a factor of four. Furthermore, Raspberry Pis are portable and many times more efficient than most CPUs, allowing them to be used in places where traditional computers can't exist (e.g. remote sensor devices).

The message passing interface (MPI) is a common paradigm for parallel distributed applications on supercomputers and was studied on \arm-based clusters. See Section~\ref{sec:related:work}. An alternative to MPI and the message passing paradigm is the tasked-based paradigm, also called asynchronous many-task (AMT). One example for AMT is the C++ standard library for concurrency and parallelism (HPX)~\cite{Heller2017}, but there are many others~\cite{chamberlain07parallelprogrammability,cilk++,carteredwards:2014:kok:2841458.2841785}. A comparative review is given in~\cite{thoman2018taxonomy}.

This study uses the three most recent Raspberry Pi models (Model 3 B, Model 3 B+, and Model 4) for a set of various benchmarks: a stencil-based one-dimensional heat equation solver (for distributed memory analysis), a stencil-based two-dimension Jacobi method solver (for shared memory and vectorization analysis), and the alternating least square (ALS) algorithm. First, the scaling on shared and distributed memory is investigated. Second, the energy consumption is compared against one of the conventional x86 architecture.


\subsection{Our Contribution}

In this paper, we present the first overview of porting and evaluating an AMT (HPX) to Raspberry Pis. We investigate the performances on all fronts, i.e. vectorization, shared memory, and distributed memory. We further investigate the distributed machine learning potential of Raspberry Pis using Phylanx, a distributed array processing toolkit that utilizes HPX for distributed memory tasks. We conclude the paper with the energy consumption benefits of Raspberry Pis.

\section{Related Work}
\label{sec:related:work}



Works using cloud training and Pis for evaluating neural nets~\cite{sajjad17raspberry} show that Raspberry Pis can be very effective. Raspberry Pis are also efficient at pre-processing data that can later be fed to a machine learning model. Collecting and pre-processing images on Raspberry Pis has been proven useful by Wang et al~\cite{wang2017detection}. Raspberry Pis are also helpful for scenarios where portability is the key. For instance, machine learning applications within a vehicle~\cite{moon2017drowsy}, at face detection and tracking~\cite{tripathy2014real}, and in botany~\cite{wani2017appropriate}. Other similar works taking advantage of Raspberry Pis have also been explored~\cite{senthilkumar2014embedded, john15alow, xu2016raspberry, baby2017smart, vzidek2017machine, tabbakha2017indoor, sarangdhar2017machine}.

While we could not find any papers on porting HPC applications to the Raspberry Pi, there many ports of HPC applications to the {\arm} ecosystem. S. McIntosh-Smith et al~\cite{McIntoshSmith2018ComparativeBO,mcintosh2019scaling} were among the first to measure performance on mainstream {\arm} HPC systems. Later, Jackson et al~\cite{10.1145/3324989.3325722} investigated the performance of distributed memory communications (MPI) through a benchmark suite utilizing MPI on ThunderX2. The energy consumption of these {\arm} processors has been investigated by the Mont-Blanc project~\cite{banchelli2019mb3}.

\section{Tools}
\label{sec:tools}
\subsection{HPX}
\begin{figure}[htpb]
    \centering
    \includegraphics[width=3.4in]{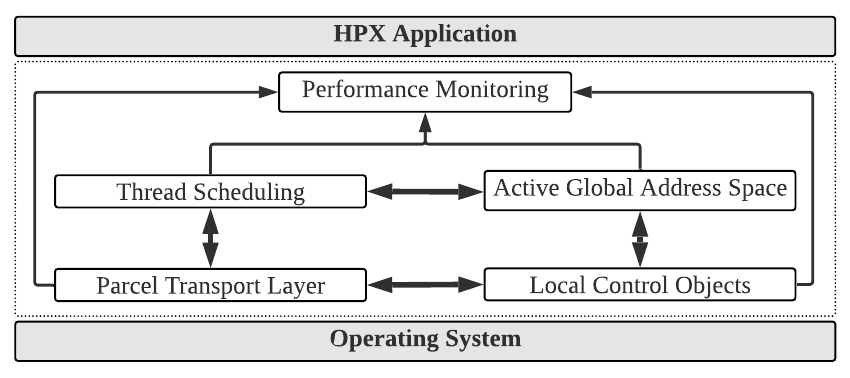}
    \caption{The architecture of HPX: Tasks, or HPX-threads, are run on top of operating system threads. A global view of the application is made possible using Active Global Address Space (AGAS). The Network Layer manages communication between tasks on different nodes.}
    \label{fig:hpx_architecture}
\end{figure}

HPX~\cite{Kaiser2020,kaiser2014hpx,Heller2017,hartmut_kaiser_2020_3675272} is based on the theoretical ParalleX~\cite{kaiser2009parallex} execution model. HPX is an asynchronous many-task runtime system with an API that closely adheres to the ISO C++ standard that enables wait-free asynchronous parallel programming including futures, channels and other synchronization primitives. Conforming with the C++ standard makes it possible for HPX code to be deployed on virtually any machine. Figure~\ref{fig:hpx_architecture} shows the architecture of HPX.

Tasks in HPX, also called \textit{HPX-threads}, are lightweight threads that are scheduled on top of the underlying operating system threads. Asynchronous execution in HPX is achieved through \textit{futures}~\cite{baker1977future} (which are placeholders for the result of a computation that has not yet been completed), \textit{continuations} (a function which is invoked when a future becomes ready), and \textit{dataflow} (which are functions that don't begin execution until their arguments are ready). These mechanisms enable HPX programmers to write fully asynchronous code. The tasks generated by the application are synchronized by Local Control Objects (LCOs), which are a family of synchronization primitives. 

The parcel~\cite{wagle2018methodology} subsystem is an active-message networking layer that ships functions to the objects they operate on. Because HPX is designed for use with distributed memory, HPX utilizes Active Global Address Space (AGAS) to track remote objects. This is achieved by assigning each object a unique Global Identifier (GID) that persists till object destruction.

A performance counter layer sits on top of these four subsystems, providing feedback to developers.

In the context of this work, HPX is used as a backend for Phylanx that is described in section~\ref{sec:phylanx}.

\subsection{Phylanx} \label{sec:phylanx} 
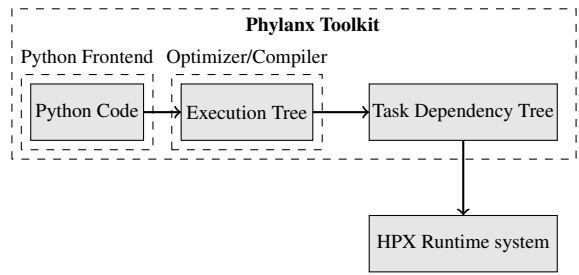
\begin{figure}[htbp]
    \centering
\begin{tikzpicture}[every node/.style={scale=0.75}]
\draw[fill=gray!20] (0,0) rectangle ++(1.5,0.75) node[pos=.5] {Python Code};
\draw[dashed] (0-0.125,0-0.125) rectangle ++(1.5+2*0.125,0.75+2*0.125);
\node[above] at (0.75,0.75+1*0.125) {Python Frontend};
\draw[->,thick] (1.5,0.75/2) -- (2,0.75/2);
\draw[fill=gray!20] (2,0) rectangle ++(1.75,0.75) node[pos=.5] {Execution Tree};
\draw[dashed] (2-0.125,0-0.125) rectangle ++(1.75+2*0.125,0.75+2*0.125);
\node[above] at (5.75/2,0.75+1*0.125) {Optimizer/Compiler};
\draw[->,thick] (3.75,0.75/2) -- (4.5,0.75/2);
\draw[fill=gray!20] (4.5,0) rectangle ++(2.5,0.75) node[pos=.5] {Task Dependency Tree};
\draw[dashed] (0-0.25,0-0.25) rectangle ++(7.5,2) ;
\node[below] at (3.75,1.725) {\textbf{Phylanx Toolkit}};
\draw[fill=gray!20] (4.5,-1) rectangle ++(2.5,-0.75) node[pos=.5] {HPX Runtime system};
\draw[->,thick] (5.75,0) -- (5.75,-1);
\end{tikzpicture}
    \caption{The architecture of Phylanx: Python code is compiled into a tree of phylanx objects called the execution tree. The execution tree is traversed, resulting in a task graph of \textit{futures} which are executed by HPX.}
    \label{fig:phylanx_arch}
\end{figure}

Phylanx~\cite{tohid2018asynchronous} is an HPX-based distributed array processing toolkit. The architecture of Phylanx is shown
in Fig.~\ref{fig:phylanx_arch}.

Phylanx provides a function decorator which access the function's abstract syntax tree and reinterprets it. While Phylanx supports a substantial subset of Python functionality, it is not intended to be a substitute for the Python interpreter. Instead, it is only a means to allow the analysis and evaluation mathematical kernels.

Phylanx works by translating Python code into a tree of phylanx
objects (\textit{primitives}) connected by futures known as the execution tree. Each phylanx
\textit{primitive} waits for its input futures to become ready before executing, then sets a future to convey its result to the next primitive(s). 



Phylanx uses an intermediate language called Phylanx Specification Language (PhySL) which is similar to Lisp. Programs can be written in Python or in this intermediate language. 

\section{Benchmarks}
\label{sec:benchmark}

This section explains the benchmarks and the parameters used to run them. Furthermore, we give a brief description of the system configuration and software versions.

\subsection{2D Jacobi Solver (Shared Memory)}

For shared memory performance, we implement a 2D stencil based on the Jacobi method.
To fully exploit the Neon pipeline, we add explicit
vectorization. As vectorizing a standard grid layout of a stencil is non-trivial, we changed the data layout to Virtual Node Scheme~\cite{boyle2015grid}. The changed data layout allows us to trivially vectorize the code in a cache friendly manner. We do not change the data layout for non-vectorized code to test GCC's ability to autovectorize a non-trivial layout.

Stencil codes are known to be memory bound due to their low Arithmetic Intensity (AI). Therefore, we do not expect many-fold increase in performance by utilizing explicit vectorization. Furthermore, we expect the compiler to auto vectorize the base code, which may limit the additional benefits of explicit vectorization.

The application is tested using strong scaling.
All benchmarks are run with a grid
size of 4096$\times$4096 iterating over a 100 time steps. The benchmark achieves
parallelism by dividing the grid into smaller grids of sizes 4096$\times y$ and
working on the smaller grids in parallel. Therefore, the number 4096 is chosen
such that three rows of the grid fits in the caches. This reduces
the number of memory transfers per iteration to three. The arithmetic intensity
for such a scenario is given by $\sfrac{1}{24}$ for double precision, and $\sfrac{1}{12}$ for
single precision respectively. Using the principles of the roofline model~\cite{williams2008roofline}, one can now calculate the optimal performance
using the following formula:

\begin{equation}
    \label{eq:1}
    P_{optimal} = Memory\ Bandwidth \times AI
\end{equation}

We use the STREAM benchmark~\cite{McCalpin1995} to compute the memory bandwidth. All performance numbers are provided in Million Lattice site Updates Per Second (MLUPs/s). For our case, 1 lattice site update corresponds to 4 floating point operations. Therefore, one can convert the performance from MLUPs/s to FLOPs/s by multiplying by 4.

\subsection{ALS}

ALS or ``Alternating Least Squares'' is a matrix factorization algorithm widely employed for
computing recommendations in streaming services, online stores, etc.

The idea behind ALS is that we factor a ratings matrix $R_{ij}$ where $i$ runs over users and $j$ runs over items, into a user matrix $U_{ik}$ and an item matrix, $V_{kj}$, and the $k$ values represent ``latent factors.''

The word \textit{alternating} is in the name because the solver alternately minimizes the least squares error arising from the values in $U$ and $V$.

For our benchmark, we use the first two hundred thousand lines of the MovieLens 20m database\footnote{\url{http://dx.doi.org/10.17632/n6sjkpy87f.5}}

\subsection{1D Heat Equation solver (Distributed memory)}

For distributed memory performance, we use HPX's optimized in-house stencil benchmark. This benchmark uses a single partition per locality, exchanges data from a single cell during halo exchange, and the exchange is made asynchronously via dataflow (allowing overlap of computation and communication).

For benchmarking, we use a heat transfer coefficient of $k=0.5$, a time step of $dt=1$, and a
grid spacing of $dx=1$. For strong scaling, we use 30
million and 60 million stencil points iterating for 100 or 500 time steps. For
weak scaling, we start with 30 million stencil points and another 30 million
stencil points are added for each node. The benchmark iterates over 100 time
steps.

\subsection{System Setup}

The CPU models Cortex-A53 and Cortex-A72 are capable of running both 32-bit (armv7l) and 64-bit (aarch64) instruction sets. On the Raspberry Pi model 3, there is no benefit in using a 64-bit operating system since there is only 1GB of memory available. Also, a 64-bit application is going to use more memory simply because variables take up more space---although this increase in memory usage is negligible in most cases. On the other hand, the Raspberry Pi 4 potentially has more memory and could benefit from a 64-bit operating system. To compensate the memory limitation imposed by hardware, all installations include 8GB swap space (which was needed by the gnu C++ compiler to compile HPX and Phylanx).

The 64-bit version of the official operating system from the Raspberry Pi Foundation is still in development. So far, the foundation has only released the 32-bit version. However, there are third party distributions which provide both versions, but many on those lack support for relatively newer Raspberry Pi 4. In our search, we found that Ubuntu Server 2020 is the only distribution that supports all three versions of boards, has both 32-bit and 64-bit release, and has a relatively up-to-date development library.

We understand that the Ubuntu distribution is an unusual choice for our experiment. Ubuntu, while very popular among developers and Linux enthusiasts, is rarely used in an HPC environment. As a result, many default settings are tuned toward desktop usage. By tweaking those settings, the desired performance can be achieved. For example, the default CPU profile is set to ``\emph{ondemand},'' which means the Linux kernel will keep the CPU frequency at lowest level allowed by hardware and only increase the frequency in response to system load. Although this setting may be acceptable for a single user, it could have significant impact on application performance, especially at startup. Throughout these tests, the CPU frequency is always kept at highest level allowed by hardware (using the CPU profile ``\emph{performances}'').

Although the {\arm} processors are praised for their low power consumption, they are not entirely immune to the problem of power consumption and heat generation. Like any other contemporary microprocessor, the power consumption happens during the switching time in logic gates. Therefore, the amount of heat produced is directly related to CPU frequency. Raspberry Pis do not have an active cooling system. To avoid any damage to the processor, the firmware will reduce the CPU frequency to reduce amount of heat production when a certain high temperature reached. The process is called \emph{thermal throttling} and it could cause inconsistency in the performance results. To avoid thermal throttling, a small aluminium heat-sink was installed on all processors and boards were transferred to our data-center which as well regulated air temperature and better air flow.

Distributed computing on the Raspberry Pi board isn't without challenges, since the performance of these distributed applications is heavily influenced by network speed and latency. Although all model B Raspberry Pis provide an Ethernet interface, the earlier models lack a dedicated network controller. Raspberry Pis model 3B and 3B+ use the USB controller to provide Ethernet interface, therefore the performance network interface is bound to the performance of USB interface. On the Raspberry Pi 3B, the network interface works at a speed of 100 Mb/s. On the Raspberry Pi model 3B+ the network interface establishes a 1 Gb/s link, but the underlying USB cannot transfer the data faster than 300 Mb/s, therefore, the actual data transfer rate is limited to that speed. Only on the Raspberry Pi Model 4B is this issue addressed, and a dedicated network controller capable of transferring data at the speed of 1 Gb/s is installed on the board.

\begin{table}[tbp]
  \centering
  \caption{Specification/Architecture of the three nodes utilised in the benchmarks.}
  \label{tab:nodes}
  \begin{tabular}{l|lll}
    \toprule
    \textbf{Model}              & Raspberry Pi 3B & Raspberry Pi 3B+ & Raspberry Pi 4B    \\ \hline
    \textbf{Micro-architecture}  & {\arm} v8-A & {\arm} v8-A & {\arm} v8   \\ \hline
    \textbf{Processor Model}  & Cortex-A53 & Cortex-A53 &  Cortex-A72   \\ \hline
    \textbf{Number of CPUs}     & 1 & 1   & 1         \\ \hline
    \textbf{Cores per CPU}    & 4 & 4   & 4      \\ \hline
    \textbf{Total Cores}       & 4 & 4 & 4            \\ \hline
    \textbf{Frequency}        & 1.2GHz & 1.4GHz & 1.5GHz    \\ \hline
    \textbf{Memory}           & 1GB & 1GB & 4GB     \\ \bottomrule
  \end{tabular}
\end{table}

\begin{table}[tbp]
    \centering
    \begin{tabular}{ll|ll}
    \toprule
     Operating System  & Ubuntu \num{20.04} LTS &  Kernel       & \num{5.4}   \\ & for {\arm}  &
     blaze\footnote{\url{https://bitbucket.org/blaze-lib/blaze}}     & 75179e6  \\
     Compilers  & gcc \num{9.3}\num{.1} &
     boost      & \num{1.71} \\
     hwloc      & \num{2.1}.\num{0} &
     gperftools & \num{2.7} \\
     lapack & \num{3.8} &
     HPX\footnote{\url{https://github.com/STEllAR-GROUP/hpx}}       & 5b9de48ab1  \\\bottomrule
    \end{tabular}
    \caption{Overview of the compilers, software, and operating system used.}
    \label{tab:software}
\end{table}

Table~\ref{tab:software} shows the libraries and compiler used to
build HPX, Phylanx, and the benchmarks. Table~\ref{tab:nodes}
specifies the hardware. For the distributed
memory benchmark, we used the network latency hiding 1D stencil 8
benchmark from HPX's in-house benchmark suite. For the shared memory benchmark, we
wrote a 2D stencil Jacobi solver with explicit vectorization. For Phylanx, the
in-house benchmark ALS Python script was used. 

\section{Results}
\label{sec:results}

\subsection{2D stencil (Shared Memory)}
\label{subsec:2d}

\begin{figure}[htbp]
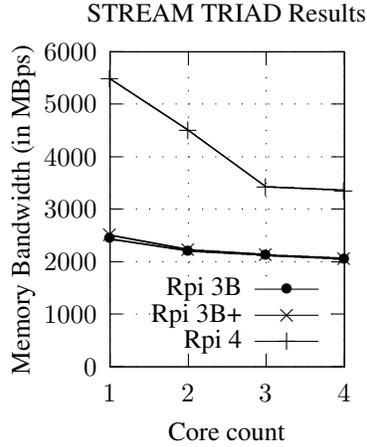

    \centering
    \captionsetup{justification=centering}
    \begin{gnuplot}[terminal=latex]
        set terminal latex rotate size 2in,2.5in
        set ylabel 'Memory Bandwidth (in MBps)'
        set xlabel 'Core count'
        set yrange [0:6000]
        set grid ytics lt 2 lc rgb "#bbbbbb"
        set grid xtics lt 2 lc rgb "#bbbbbb"
        set title 'STREAM TRIAD Results'
        set key bottom right
        
        set xtics (1, 2, 3, 4)
        
        plot \
        'plots/2d_stencil/rpi3/bandwidth' using 1:2 title 'Rpi 3B' lw 1 lt 7 lc -1 w lp, \
        'plots/2d_stencil/rpi3p/bandwidth' using 1:2 title 'Rpi 3B+' lw 1 lt 9 lc -1 w lp, \
        'plots/2d_stencil/rpi4/bandwidth' using 1:2 title 'Rpi 4' lw 1 lt 1 lc -1 w lp
    \end{gnuplot}
    \caption{Memory Bandwidth results using the STREAM TRIAD Benchmark with an array size of 10M elements}
    \label{fig:stream}
\end{figure}

Figure~\ref{fig:stream} depicts the STREAM TRIAD results. It can be seen that the Raspberry Pi 3B and 3B+ have very low memory bandwidth. The memory bandwidth is almost fully saturated by a single PU, and decreases with the number of PUs. For the Raspberry Pi 4, we observe a similar behavior. Memory bandwidth is expected to increase or be the same (if saturation is achieved) on a single NUMA domain. Here we see a sharp decline as we increase the number of PUs. The memory bus and controllers can only handle a certain amount of memory bandwidth and concurrency at the same time. This can be a possible reason behind the decline.

\begin{figure}[htpb]
    \centering
    \captionsetup{justification=centering}
    \begin{gnuplot}[terminal=latex]
        set terminal latex rotate size 3.5in,2.8in
        set ylabel 'Performance (in MLUPs/s)' 
        set xlabel 'Core count'
        
        set size 1,1
        set multiplot layout 2,2 title '2D Stencil: Raspberry Pi 4'
        set grid ytics lt 2 lc rgb "#bbbbbb"
        set grid xtics lt 2 lc rgb "#bbbbbb"

        unset key
        
        set size 0.52,0.8
        set origin 0,0.1
        
        set title 'Single Precision'
        
        set xtics (1, 2, 3, 4)
        
        set yrange [200:700]
        
        plot \
        'plots/2d_stencil/rpi4/rpi4_float' using 1:($3) title 'scalar' lw 1 lc -1 w lp, \
        'plots/2d_stencil/rpi4/rpi4_nfloat' using 1:($3) title 'vector' lw 1 lt 7 lc -1 w lp, \
        'plots/2d_stencil/rpi4/bandwidth' using 1:($2/8) title 'Expected peak Max' lw 2 lt 9 lc -1 w lp, \
        
        set size 0.48,0.8
        set origin 0.52,0.1
        
        set title 'Double Precision'
        
        set yrange [100:350]
        
        unset ylabel
        
        plot \
        'plots/2d_stencil/rpi4/rpi4_double' using 1:($3) title 'scalar' lw 1 lc -1 w lp, \
        'plots/2d_stencil/rpi4/rpi4_ndouble' using 1:($3) title 'vector' lw 1 lt 7 lc -1 w lp, \
        'plots/2d_stencil/rpi4/bandwidth' using 1:($2/16) title 'Expected peak Max' lw 1 lt 9 lc -1 w lp, \
        
        set size 0.9,0.21
        set origin 0.1,-0.07
        unset title
        set key horizontal center left
        unset tics
        unset xlabel
        unset ylabel
        set yrange [0:1]
        plot 2 t 'scalar' lw 2 lc -1 w lp, \
             2 t 'vector' lw 2 lt 7 lc -1 w lp, \
             2 t 'Expected Peak' lw 2 lt 9 lc -1 w lp, \
        
        unset multiplot
    \end{gnuplot}
    \caption{2D stencil (Raspberry Pi 4): Grid size of 4096$\times$4096 iterated over a 100 time steps.}
\end{figure}

\begin{figure}[tb]
    \centering
    \captionsetup{justification=centering}
    \begin{gnuplot}[terminal=latex]
        set terminal latex rotate size 3.5in,2.8in
        set ylabel 'Performance (in MLUPs/s)' 
        set xlabel 'Core count'
        
        set size 1,1
        set multiplot layout 2,2 title '2D Stencil: Raspberry Pi 3B+'
        set grid ytics lt 2 lc rgb "#bbbbbb"
        set grid xtics lt 2 lc rgb "#bbbbbb"

        unset key
        
        set size 0.52,0.8
        set origin 0,0.1
        
        set title 'Single Precision'
        
        set xtics (1, 2, 3, 4)
        
        plot \
        'plots/2d_stencil/rpi3p/rpi3p_float' using 1:($3) title 'scalar' lw 1 lc -1 w lp, \
        'plots/2d_stencil/rpi3p/rpi3p_nfloat' using 1:($3) title 'vector' lw 1 lt 7 lc -1 w lp, \
        'plots/2d_stencil/rpi3p/bandwidth' using 1:($2/8) title 'Expected peak Max' lw 2 lt 9 lc -1 w lp, \
        
        set size 0.48,0.8
        set origin 0.52,0.1
        
        set title 'Double Precision'
        
        unset ylabel
        
        plot \
        'plots/2d_stencil/rpi3p/rpi3p_double' using 1:($3) title 'scalar' lw 1 lc -1 w lp, \
        'plots/2d_stencil/rpi3p/rpi3p_ndouble' using 1:($3) title 'vector' lw 1 lt 7 lc -1 w lp, \
        'plots/2d_stencil/rpi3p/bandwidth' using 1:($2/16) title 'Expected peak Max' lw 1 lt 9 lc -1 w lp, \
        
        set size 0.9,0.21
        set origin 0.1,-0.07
        unset title
        set key horizontal center left
        unset tics
        unset xlabel
        unset ylabel
        set yrange [0:1]
        plot 2 t 'scalar' lw 2 lc -1 w lp, \
             2 t 'vector' lw 2 lt 7 lc -1 w lp, \
             2 t 'Expected Peak' lw 2 lt 9 lc -1 w lp, \
        
        unset multiplot
    \end{gnuplot}
    \caption{2D stencil (Raspberry Pi 3B+): Grid size of 4096$\times$4096 iterated over a 100 time steps.}
\end{figure}

Using Equation~\ref{eq:1}, we find the expected peak performance. We use the Linux \texttt{perf} utility to retrieve performance counters. The Gnu compiler was readily able to auto vectorize our 2D stencil with minimal differences in instruction counts. Visible improvements, however, were seen in cache-references and cache-misses.
This means that the approach taken by GCC to auto vectorize the code differs in data layout.

For the Raspberry Pi 4, the best recorded performance is for a core count of 2 and 3. This is because the memory bandwidth decreases with the core count. Our measurements are on par with the expected peak performance.

Table~\ref{tab:perf_rpi4} shows the hardware counter values we measured. Our \texttt{perf} results show similar instruction counts, but about 50\% less cache misses for explicitly vectorized codes. This shows that our Virtual Node Scheme~\cite{boyle2015grid} is more cache-friendly compared to the autovectorization layout utilized by GCC.

\begin{table}[h!]
    \centering
    \caption{Hardware counters for Raspberry Pi4}
    \begin{tabular}{lcc}
    \toprule
     Data Type  & Instructions & Cache-Misses \\\midrule
     float  & 6,168,850,721 & 288,165,018   \\
     nsimd float & 5,858,534,460 & 210,042,447  \\
     double & 11,553,460,548 & 641,066,436 \\
     nsimd double & 11,147,560,795 & 411,352,041 \\\bottomrule
    \end{tabular}
    \label{tab:perf_rpi4}
\end{table}

\begin{figure}[b]
    \centering
    \captionsetup{justification=centering}
    \begin{gnuplot}[terminal=latex]
        set terminal latex rotate size 3.5in,2.8in
        set ylabel 'Performance (in MLUPs/s)' 
        set xlabel 'Core count'
        
        set size 1,1
        set multiplot layout 2,2 title '2D Stencil: Raspberry Pi 3B'
        set grid ytics lt 2 lc rgb "#bbbbbb"
        set grid xtics lt 2 lc rgb "#bbbbbb"

        unset key
        
        set size 0.52,0.8
        set origin 0,0.1
        
        set title 'Single Precision'
        
        set xtics (1, 2, 3, 4)
        
        plot \
        'plots/2d_stencil/rpi3/rpi3_float' using 1:($3) title 'scalar' lw 1 lc -1 w lp, \
        'plots/2d_stencil/rpi3/rpi3_nfloat' using 1:($3) title 'vector' lw 1 lt 7 lc -1 w lp, \
        'plots/2d_stencil/rpi3/bandwidth' using 1:($2/8) title 'Expected peak Max' lw 2 lt 9 lc -1 w lp, \
        
        set size 0.48,0.8
        set origin 0.52,0.1
        
        set title 'Double Precision'
        
        unset ylabel
        
        plot \
        'plots/2d_stencil/rpi3/rpi3_double' using 1:($3) title 'scalar' lw 1 lc -1 w lp, \
        'plots/2d_stencil/rpi3/rpi3_ndouble' using 1:($3) title 'vector' lw 1 lt 7 lc -1 w lp, \
        'plots/2d_stencil/rpi3/bandwidth' using 1:($2/16) title 'Expected peak Max' lw 1 lt 9 lc -1 w lp, \
        
        set size 0.9,0.21
        set origin 0.1,-0.07
        unset title
        set key horizontal center left
        unset tics
        unset xlabel
        unset ylabel
        set yrange [0:1]
        plot 2 t 'scalar' lw 2 lc -1 w lp, \
             2 t 'vector' lw 2 lt 7 lc -1 w lp, \
             2 t 'Expected Peak' lw 2 lt 9 lc -1 w lp, \
        
        unset multiplot
    \end{gnuplot}
    \caption{2D stencil (Raspberry Pi 3B): Grid size of 4096$\times$4096 iterated over a 100 time steps}
\end{figure}

For the Raspberry Pi 3B+ and 3B, we see very similar performance. This is because the two models differ only in the clock speeds. The Raspberry Pi 3B clocks at 1.2GHz whereas 3B+ clocks at 1.4GHz. This allows for a 16\% improvement in performance, which is noticeable at lower thread counts. At a core count of 1 and 2, 3B+ performs noticeably better than 3B. At higher core counts, the performance is almost the same. For both these processors, we are not able to achieve the expected peak performance. Explicit vectorization is not able to help boost the performance either. While 1 core performance can be attributed to the limited compute, many core performance is sub par. Table~\ref{tab:perf_rpi3p} describes the major contributing hardware counters when all four cores were used. From the instruction count, it is clear that GCC does well at auto vectorization. However, similar to the Raspberry Pi 4, GCC fails to exploit the data layout resulting in a higher cache-miss count. While Raspberry Pis do not allow access to hardware stall counters, we believe that explicit vectorization gains from having lower number of memory transactions in-flight. This explains why the explicitly vectorized results are sub par as well. A redesign of vectorized elements should help alleviate the problems given the small cache size and higher instruction count.

\begin{table}[h!]
    \centering
    \caption{Hardware counters for Raspberry Pi3B+}
    \begin{tabular}{lcc}
    \toprule
     Data Type  & Instructions & Cache-Misses \\\midrule
     float  & 8,973,631,540 & 147,993,989   \\
     nsimd float & 13,484,044,923 & 133,391,170  \\
     double & 22,009,615,376 & 296,496,230 \\
     nsimd double & 23,158,335,089 & 279,417,800 \\\bottomrule
    \end{tabular}
    \label{tab:perf_rpi3p}
\end{table}

\subsection{ALS}

For the ALS benchmaark, the performance of the Pi4 is nearly double the value obtained for the Pi3 and Pi3+, despite the modest difference in clock speed.

\begin{figure}[htpb]
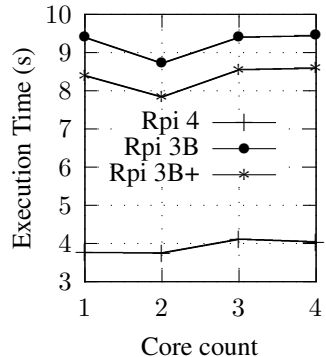

    \centering
    \captionsetup{justification=centering}
    \begin{gnuplot}[terminal=latex]
        set terminal latex rotate size 1.85in,2.25in
        set ylabel 'Execution Time (s)'
        set xlabel 'Core count'
        set datafile separator ","
        set grid ytics lt 2 lc rgb "#bbbbbb"
        set grid xtics lt 2 lc rgb "#bbbbbb"

        set key center
        set title 'Performance of the ALS Benchmark'
        set xtics (1, 2, 3, 4)
        
        plot \
        'plots/als/als_rpi4.dat' using 1:2 title 'Rpi 4' lw 1 lc -1 w lp, \
        'plots/als/als_rpi3.dat' using 1:2 title 'Rpi 3B' lw 1 lt 7 lc -1 w lp, \
        'plots/als/als_rpi3p.dat' using 1:2 title 'Rpi 3B+' lw 1 lt 9 lc -1 w lp
    \end{gnuplot}
    \caption{Here we see the performance of the ALS benchark on the Pi3, the Pi3B+ and the Pi4. The Pi4 provides almost twice the performance.}
    \label{fig:benchamrk:phylanx:als}
\end{figure}

Because this application, as it is written, is fairly memory intensive, the larger memory bandwidth and larger L2 cache of the Pi4 ($1$mb vs. $512$kb) are the significant factors here.

Our ALS application receives little benefit from parallelism. We believe that, with future effort at improving the code, we will be able to get better parallel speedup.

\subsection{1D stencil (Distributed Memory)}

\begin{figure}[tb]
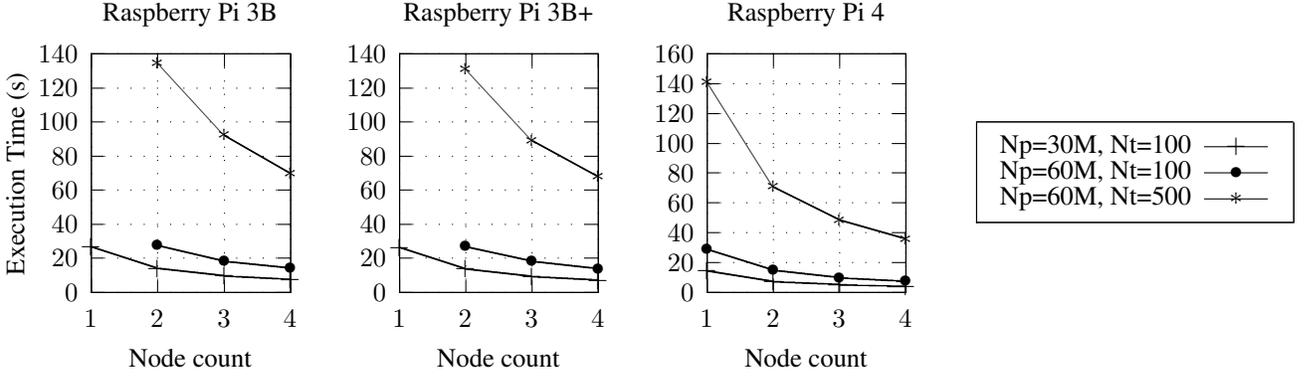
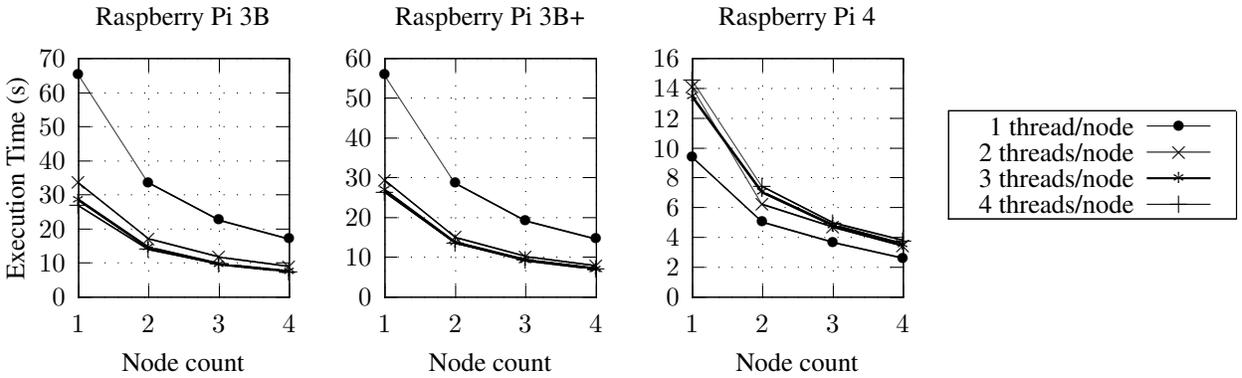

    \centering
    \begin{subfigure}{\linewidth}
        \centering
        \begin{gnuplot}[terminal=latex]
            set terminal latex rotate size 7in,2.1in
            
            set ylabel "Execution Time (s)"
            set xlabel "Node count"
            
            set size 1,1
            set multiplot layout 1,4
            set grid ytics lt 2 lc rgb "#bbbbbb"
            set grid xtics lt 2 lc rgb "#bbbbbb"
    
            unset key
            
            set size 0.25,1
            set origin 0,0
    
            set title 'Raspberry Pi 3B'
            
            set xtics (1, 2, 3, 4)
            
            plot \
            'plots/1d_stencil/rpi3/strong_30m_100.dat' using 1:2 title 'Rpi 3B' lw 1 lc -1 w lp, \
            'plots/1d_stencil/rpi3/strong_60m_100.dat' using 1:2 title 'Rpi 4' lw 1 lt 7 lc -1 w lp, \
            'plots/1d_stencil/rpi3/strong_60m_500.dat' using 1:2 title 'Rpi 4' lw 1 lt 9 lc -1 w lp
            
            set size 0.23,1
            set origin 0.25,0
            
            unset ylabel
    
            set title 'Raspberry Pi 3B+'
            set xtics (1, 2, 3, 4)
            
            plot \
            'plots/1d_stencil/rpi3p/strong_30m_100' using 1:2 title 'Rpi 3B' lw 1 lc -1 w lp, \
            'plots/1d_stencil/rpi3p/strong_60m_100' using 1:2 title 'Rpi 4' lw 1 lt 7 lc -1 w lp, \
            'plots/1d_stencil/rpi3p/strong_60m_500' using 1:2 title 'Rpi 4' lw 1 lt 9 lc -1 w lp
    
            set size 0.23,1
            set origin 0.48,0
            
            set title 'Raspberry Pi 4'
            
            set xtics (1, 2, 3, 4)
            
            plot \
            'plots/1d_stencil/rpi4/strong_30m_100.dat' using 1:2 title 'Rpi 3B' lw 1 lc -1 w lp, \
            'plots/1d_stencil/rpi4/strong_60m_100.dat' using 1:2 title 'Rpi 4' lw 1 lt 7 lc -1 w lp, \
            'plots/1d_stencil/rpi4/strong_60m_500.dat' using 1:2 title 'Rpi 4' lw 1 lt 9 lc -1 w lp
    
            set size 0.29,0.38
            set origin 0.71,0.32
            unset title
            set key vertical center right
            unset tics
            unset xlabel
            unset ylabel
            set yrange [0:1]
            plot 2 t 'Np=30M, Nt=100' lw 2 lc -1 w lp, \
                 2 t 'Np=60M, Nt=100' lw 2 lt 7 lc -1 w lp, \
                 2 t 'Np=60M, Nt=500' lw 2 lt 9 lc -1 w lp, \

            unset multiplot
        \end{gnuplot}
        \caption{1D stencil: Strong scaling results (ran on all threads per node) for different stencil points and iterations.}
        \label{fig:strong}
    \end{subfigure}
    \begin{subfigure}{\linewidth}
        \centering
        \begin{gnuplot}[terminal=latex]
            set terminal latex rotate size 6.7in,2.1in
            
            set ylabel "Execution Time (s)"
            set xlabel "Node count"
            
            set size 1,1
            set multiplot layout 1,4
            set grid ytics lt 2 lc rgb "#bbbbbb"
            set grid xtics lt 2 lc rgb "#bbbbbb"
    
            unset key
            
            set size 0.26,1
            set origin 0,0
    
            set title 'Raspberry Pi 3B'
            
            set xtics (1, 2, 3, 4)
            
            plot \
            'plots/1d_stencil/rpi3/30m_100_1' using 1:2 title 'Rpi 3B' lw 1 lt 7 lc -1 w lp, \
            'plots/1d_stencil/rpi3/30m_100_2' using 1:2 title 'Rpi 4' lw 1 lt 9 lc -1 w lp, \
            'plots/1d_stencil/rpi3/30m_100_3' using 1:2 title 'Rpi 4' lw 1 lt 11 lc -1 w lp, \
            'plots/1d_stencil/rpi3/30m_100_4' using 1:2 title 'Rpi 4' lw 1 lt 1 lc -1 w lp
            
            set size 0.24,1
            set origin 0.26,0
            
            unset ylabel
    
            set title 'Raspberry Pi 3B+'
            set xtics (1, 2, 3, 4)
            
            set yrange [0:60]
            
            plot \
            'plots/1d_stencil/rpi3p/30m_100_1' using 1:2 title 'Rpi 3B' lw 1 lt 7 lc -1 w lp, \
            'plots/1d_stencil/rpi3p/30m_100_2' using 1:2 title 'Rpi 4' lw 1 lt 9 lc -1 w lp, \
            'plots/1d_stencil/rpi3p/30m_100_3' using 1:2 title 'Rpi 4' lw 1 lt 11 lc -1 w lp, \
            'plots/1d_stencil/rpi3p/30m_100_4' using 1:2 title 'Rpi 4' lw 1 lt 1 lc -1 w lp
    
            set size 0.24,1
            set origin 0.50,0
            
            set title 'Raspberry Pi 4'
            
            set xtics (1, 2, 3, 4)
            set yrange [0:16]
            
            plot \
            'plots/1d_stencil/rpi4/30m_100_1' using 1:2 title 'Rpi 3B' lw 1 lt 7 lc -1 w lp, \
            'plots/1d_stencil/rpi4/30m_100_2' using 1:2 title 'Rpi 4' lw 1 lt 9 lc -1 w lp, \
            'plots/1d_stencil/rpi4/30m_100_3' using 1:2 title 'Rpi 4' lw 1 lt 11 lc -1 w lp, \
            'plots/1d_stencil/rpi4/30m_100_4' using 1:2 title 'Rpi 4' lw 1 lt 1 lc -1 w lp
    
            set size 0.26,0.4
            set origin 0.74,0.34
            unset title
            set key vertical center right
            unset tics
            unset xlabel
            unset ylabel
            set yrange [0:1]
            plot 2 t '1 thread/node' lt 7 lc -1 w lp, \
                 2 t '2 threads/node' lt 9 lc -1 w lp, \
                 2 t '3 threads/node' lt 11 lc -1 w lp, \
                 2 t '4 threads/node' lt 1 lc 7 w lp

            unset multiplot
        \end{gnuplot}
        \caption{1D stencil: Strong scaling results (ran on all threads per node) for 30 million stencil points iterated over a 100 time steps.}
        \label{fig:threads}
    \end{subfigure}
    \caption{1D stencil: Strong Scaling results for different combinations of stencil points and iterations, and threads per node.}
    \label{fig:1d_strong}
\end{figure}

\begin{figure}[tpb]
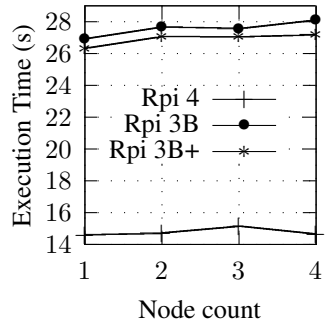

    \centering
    \captionsetup{justification=centering}
    \begin{gnuplot}[terminal=latex]
        set terminal latex rotate size 1.85in,2.1in
        set ylabel 'Execution Time (s)'
        set xlabel 'Node count'
        
        set grid ytics lt 2 lc rgb "#bbbbbb"
        set grid xtics lt 2 lc rgb "#bbbbbb"

        set key center
        set title '1D stencil: Weak Scaling (All threads)'
        set yrange [14:29]
        set xtics (1, 2, 3, 4)
        
        plot \
        'plots/1d_stencil/rpi4/weak_30m_100' using 1:2 title 'Rpi 4' lw 1 lc -1 w lp, \
        'plots/1d_stencil/rpi3/weak_30m_100' using 1:2 title 'Rpi 3B' lw 1 lt 7 lc -1 w lp, \
        'plots/1d_stencil/rpi3p/weak_30m_100' using 1:2 title 'Rpi 3B+' lw 1 lt 9 lc -1 w lp
    \end{gnuplot}
    \caption{1D stencil: weak scaling results (ran on all threads per node) for 30M stencil points iterated over 100 time steps.}
\end{figure}

Strong scaling results are provided in Figure~\ref{fig:1d_strong}. The benchmark was designed to hide latency and allow for an almost linear scaling. One interesting point to note is that the benchmark fails to run with 60 million stencil point configuration on Raspberry Pi 3B and 3B+ because memory is insufficient.

We see a behavior similar to what was observed to the 2D stencil (Sec.~\ref{subsec:2d}) while observing strong scaling for different threads/node. For the Raspberry Pi 4, single threaded runs resulted in the best performance, and all other multi-threaded variations resulted in a similar lower performance. For the Raspberry Pi 3B and 3B+, we see improvements in result as we progress to all threads per node. A significant difference can be observed going from single core to two cores per node. Moving from two to three cores provide an additional 15\% improvement. No noticeable difference is observed thereafter. The additional clock speed helps Raspberry Pi 3B+ to perform about 12\% faster than 3B.

For all the Raspberry Pi models, we see a slight rise in execution time going from one to many nodes. The rise in execution time is attributed to the initialization time of the communication protocol and are not a result of increased execution times of the kernel. Again, we see the Raspberry Pi 3B+ showing better performances compared to 3B due to the higher clock speed.

\section{Energy consumption}
\label{sec:energy}
One interesting aspect of this kind of hardware is the energy consumption~\cite{anwaar2015energy,Ou:2012:ECA:2310096.2310142,cloutier2016raspberry}.
We compute the cost for the simulations in US dollars for the 1D stencil benchmark using HPX and for the ALS benchmark using Phylanx.

\begin{figure}[htpb]
    \centering
    \begin{subfigure}{0.45\linewidth}
        \centering
        \includegraphics[width=0.75\linewidth]{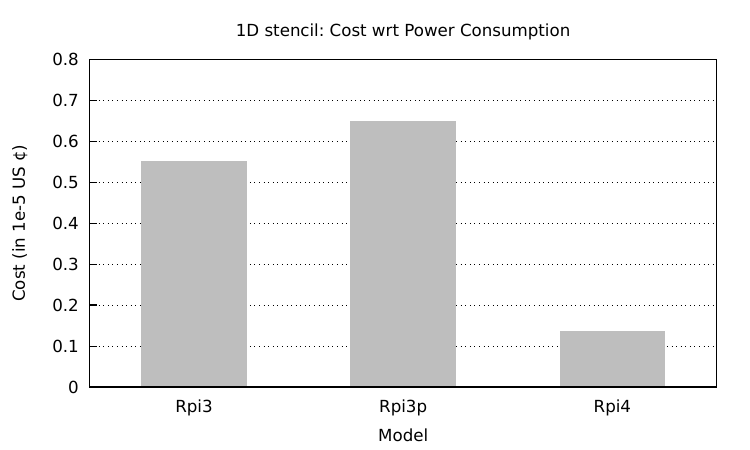}
        \caption{Cost with respect to power consumption for the 1D stencil code using 30 million stencil points per iteration and a total of 100 iterations.}
        \label{fig:energy}
    \end{subfigure}
    \hfill
    \begin{subfigure}{0.45\linewidth}
        \centering
		\includegraphics[width=0.75\linewidth]{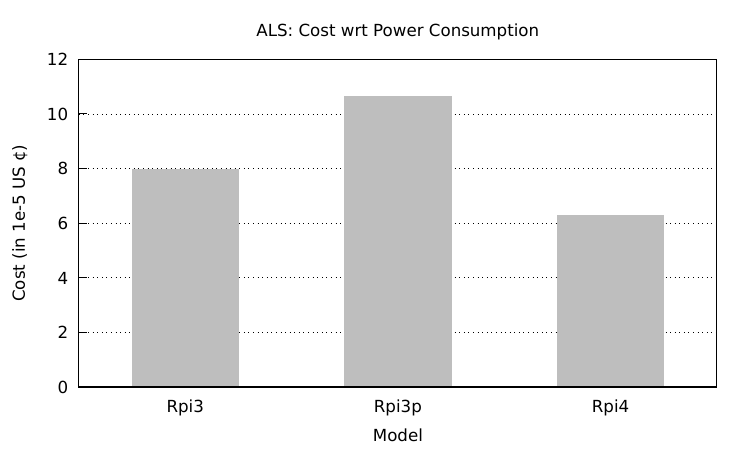}
        \caption{Cost with respect to power consumption for the ALS benchmark.}
        \label{fig:energy:als}
    \end{subfigure}
    \caption{Cost with respect to power consumption for (a) 1D stencil and (b) ALS benchmark. The power consumption for all models was obtained using the Linux command \texttt{stress} for all four cores. The costs are for the state of Louisiana, since the PI cluster was located there.}
    \label{fig:als_out}
\end{figure}

For the power consumption, we used the measurements obtained by the Linux command\footnote{\url{https://linux.die.net/man/1/stress}} \lstinline{stress --cpu 4}  in Table~\ref{tab:power} which were provided by Jeff Geerling\footnote{\url{https://www.pidramble.com/wiki/benchmarks/power-consumption}}. Note that the Raspberry Pi device does not provide a hardware-enabled power measurement, and we could not use the PAPI or APEX library to obtain these values. For the price per kWh, we use the average residential electricity
rate in Baton Rouge is \num{8.2}\textcent/kWh~\footnote{\url{https://www.electricitylocal.com/states/louisiana/baton-rouge/}}, since the hardware is located there.

\begin{table}[h!]
    \centering
    \caption{Power consumption of various Raspberry Pi models using all four cores obtained by running the Linux command \texttt{stress --cpu 4}.}
    \begin{tabular}{lcc}
    \toprule
     Model  & Watt & milli Ampere \\\midrule
     PI 3B  & 3.7 & 730   \\
     PI 3B+ & 5.1 & 980  \\
     PI 4 & 6.4 & 1280 \\\bottomrule
    \end{tabular}
    \label{tab:power}
\end{table}

First, the cost in US \textcent~per iteration for the 1D stencil using 30 million stencil points, see Figure~\ref{fig:threads} was calculated. Figure~\ref{fig:energy} shows that the cost per iteration is very low for all models, the PI 4 has the lowest cost of all, because the computation time is around five times less. The PI 3B+ has the highest cost, since the time difference to the PI 3 is close, but the power consumption is around one third more. 

Second, the cost in US \textcent~for the ALS benchmark is shown in Figure~\ref{fig:energy:als}. Again, the PI 4 has the lowest cost since the computation finished in half of the time. For the PI 3B+, the cost is the highest again for the same reason as above.

However, the trade-off for energy consumption is clearly performance, as one can
see in the previous section for comparing the computation time. This aligns with
the references in the related work section for running other parallel and
distributed application on Raspberry Pis.

\section{Conclusion and Outlook}
\label{sec:conclusion}

In this paper we have described how to use a cluster of Raspberry Pi (32-bit and 64-bit) for a number of small benchmarks that are part of the HPX and Phylanx toolkit. To support 32-bit and 64-bit options, using the Ubuntu Server 2020 operating system is the best option. However, we examined performance issues related to vectorization, threading, clock frequency settings, and operating system choice. We note that, for purposes of benchmarking, an appropriate CPU profile needs to be set instead of the default option. In our case, we used \emph{performance}.

We found that, due to the limited memory bandwidth, Pis are often unable to make use of all four cores. The STREAM TRIAD benchmark shows that memory bandwidth decreases for the Raspberry Pi 4 as threads are added. This is probably why performance peaks with 2 or 3 cores. Extending the observation to a distributed scale, we find that our 1D stencil benchmark scales almost linearly on a distributed scale while worsening as more threads are added per node.

In terms of energy consumption, we observed that for both benchmarks the Raspberry Pi 4 had the lowest power consumption and thus the lowest costs. However, clearly trade-off is the computational time for all three models. Furthermore, limited scalability was observed for the Raspberry Pi 3. In short, with some effort at configuration, a Raspberry Pi cluster can provide modest performance at a reasonable cost. 

\subsection{Outlook}
From our experience, two possible use cases of the Raspberry Pi cluster can be considered. First, for the Raspberry Pi 4 cluster, an application in teaching parallel and distributing computing is imaginable. The students can see the behavior of adding threads or multiple nodes to the performance of the application. Thus, the students do not occupy more performant nodes which might be required for other valuable research.  

Second, the Raspberry Pi devices provide interfaces to attach sensors such as temperature and humidity sensors. Raspberry Pis are used in field studies to collect sensor data and offloading them to a more powerful device to carry analysis over the data. Phylanx can help in such scenarios by processing data at the Pi's end before offloading.   

\section{Acknowledgments}
This work was partly funded by DTIC Contract FA8075-14-D-0002/0007 and the
Center of Computation \& Technology at Louisiana State University.

\section*{Copyright notice}
\textcopyright 2020 IEEE. Personal use of this material is permitted. Permission from IEEE must be obtained for all other uses, in any current or future media, including reprinting/republishing this material for advertising or promotional purposes, creating new collective works, for resale or redistribution to servers or lists, or reuse of any copyrighted component of this work in other works.

\bibliographystyle{IEEEtran}
\bibliography{references}

\end{document}